\documentclass{./aims/aims}
\usepackage{amsmath}
\usepackage{amssymb}
\usepackage{dsfont}
\usepackage{bm}
\usepackage{bbm}
  \usepackage{paralist}
  \usepackage{graphics} 
  \usepackage{epsfig} 
\usepackage{graphicx}  \usepackage{epstopdf}

\usepackage{subfig}
\usepackage{booktabs}
\usepackage{arydshln}
\usepackage[]{algorithm}
\usepackage{algpseudocode}
 \usepackage[colorlinks=true]{hyperref}
\hypersetup{urlcolor=blue, citecolor=red}

\def\currentvolume{X}
 \def\currentissue{X}
  \def\currentyear{200X}
   \def\currentmonth{XX}
    \def\ppages{X--XX}

\theoremstyle{definition}

\usepackage[pagewise]{lineno}\nolinenumbers
\title[A Log-Gaussian Cox Process with SMC for Line Narrowing] 
      {A Log-Gaussian Cox Process with Sequential Monte Carlo for Line Narrowing in Spectroscopy}

\author[T. H\"ark\"onen, E. Hannula, M. T. Moores, E. M. Vartiainen, and L. Roininen]{}


 \email{teemu.harkonen@lut.fi}
 \email{emma.hannula@student.lut.fi}
 \email{mmoores@uow.edu.au}
 \email{erik.vartiainen@lut.fi}
 \email{lassi.roininen@lut.fi}

\thanks{$^*$ Corresponding author: Teemu Härkönen}

\begin{document}
\maketitle

\centerline{\scshape Teemu Härkönen$^*$, Emma Hannula}
\medskip
{\footnotesize
 \centerline{School of Engineering Science}
 \centerline{LUT University}
   \centerline{Yliopistonkatu 34, FI-53850 Lappeenranta, Finland}
} 

\medskip

\centerline{\scshape Matthew T. Moores}
\medskip
{\footnotesize
 \centerline{National Institute for Applied Statistics Research Australia}
   \centerline{University of Wollongong}
   \centerline{Wollongong NSW 2522, Australia}
}

\medskip

\centerline{\scshape Erik M. Vartiainen, Lassi Roininen}
\medskip
{\footnotesize
 \centerline{School of Engineering Science}
 \centerline{LUT University}
   \centerline{Yliopistonkatu 34, FI-53850 Lappeenranta, Finland}
}

\bigskip

 \centerline{(Communicated by the associate editor name)}

\begin{abstract}
We propose a statistical model for narrowing line shapes in spectroscopy that are well approximated as linear combinations of Lorentzian or Voigt functions. We introduce a log-Gaussian Cox process to represent the peak locations thereby providing uncertainty quantification for the line narrowing. Bayesian formulation of the method allows for robust and explicit inclusion of prior information as probability distributions for parameters of the model. Estimation of the signal and its parameters is performed using a sequential Monte Carlo algorithm followed by an optimization step to determine the peak locations. Our method is validated using a simulation study and applied to a mineralogical Raman spectrum.
\end{abstract}

%
\section{Introduction}

In signal processing, it is often fruitful to analyze a time series through its spectral density representation, or periodogram \cite{Brockwell:2016}.
The frequencies or wavenumbers contained in the observed signal are displayed along the horizontal axis, while the corresponding energy intensities are displayed on the vertical axis.
In many cases, the spectrum exhibits peaks with a certain line shape, which can be characterized by a spectral density function.
Examples include ocean waves \cite{Stewart:2004},  ionospheric spectra measured by incoherent scatter radar \cite{virtanen:2021}, and X-ray spectra used for industrial quality control \cite{suuronen:2020}.
In this paper we focus on the specific example of Raman spectroscopy \cite{Smith:2019}, but the methods that we discuss are much more broadly applicable to many other kinds of spectra.

Line narrowing of spectral line shapes is a mathematical procedure that is used to improve the resolution of spectroscopy.
The principal aim of line narrowing is to reduce overlap of the line shapes and to infer more accurate information on the line shape position.
There is a long history of research in various approaches for line narrowing based on ideas such as Fourier self-deconvolution \cite{ Kauppinen:91, fsdarticle, kauppAsump}, Tikhonov regularization \cite{ Cui:2020, Liu:2013}, maximum entropy \cite{ Buttingsrud:2004, Fonfria:2005}, and Bayesian inference \cite{Frohling:2016,Razul:2003,Ritter:1994}.
Nevertheless, the applicability of these algorithms tends to be limited by requirements on input parameters not known in general, meaning that they must be hand-tuned, and also by the signal-to-noise ratio of the measurements.

To consider a specific example of line-narrowing algorithm, Line-shape Optimized Maximum Entropy linear Prediction (LOMEP) works well only
for spectra with a limited number of spectral lines, which all have the same \textit{a priori} known line shape \cite{ Kauppinen:91}.
Nevertheless, the fundamental idea behind LOMEP is very appealing.
It uses a technique called Fourier self-deconvolution \cite{fsdarticle,kauppAsump}, where the Fourier transform of the data is divided by the Fourier transform of a parameterized kernel function, resulting in a signal with narrower line shapes.
This requires \textit{a priori} knowledge of the line-shape function in the original spectrum.
A low-pass filter is applied to the obtained non-decaying signal to ensure high signal-to-noise ratio.
After this, a maximum-entropy, linear-prediction algorithm  is used to predict the non-decaying signal to yield new narrower line shapes while preserving their respective amplitudes.

In our previous work \cite{Harkonen:2020}, we incorporated LOMEP as a pre-processing step for empirical Bayesian inference on the line shape parameters using a Sequential Monte Carlo (SMC) algorithm \cite{DelMoral:2006}.
Here, we extend this approach to a fully-Bayesian statistical model that is capable of providing posterior distributions for the estimated line-narrowed spectrum, along with posterior distributions for the line width and impulse-response length.
We sample these model parameters and realizations using SMC, which utilizes a collection of particles to approximate the probability distributions of interest.
Our SMC algorithm  provides a scalable and parallelizable method of statistical inference for spectroscopic data.

The LOMEP algorithm is known to suffer from peak splitting, where individual peaks in the underlying signal are split into two or more peaks, thereby overestimating the number of peaks that are present \cite{Kauppinen:1992}.
The key contribution of this paper is the introduction of a Log-Gaussian Cox Process (LGCP) \cite{Moller:1998} for this specific application, as a statistical model for the peak locations.
An LGCP is a doubly-stochastic process, whose output is defined as a Poisson process and whose intensity function is modelled as a log-Gaussian process.  
This enables modelling of observed point data that exhibit clustering, the intensity of which can vary spatially or temporally.
The LGCP has previously been applied to statistical modelling of disease incidence data \cite{ Diggle:2013}, wildfire occurrences \cite{Serra:2014}, and crime incidence \cite{Shirota:2017}.
Here, the intensity function of the LGCP can be considered to provide automatic smoothing where the parameters of the smoothing kernel are inferred statistically.
For the LGCP intensity function, we use a maximum \textit{a posteriori} (MAP) estimate.
In addition to the primary interest of line narrowing, the posterior distributions for the line widths and peak locations have immediate applications in being incorporated as prior distributions for further statistical spectrum analysis techniques \cite{ Harkonen:2020, Moores:2018}. 

The remainder of this paper is structured as follows.
In Section \ref{sec:model}, we present our Bayesian statistical model for spectral measurements.
In Section \ref{sec:smc}, this is followed by a description of the SMC algorithm.
In Section \ref{sec:lgcp}, we formulate our log-Gaussian Cox process model to better estimate the underlying line-narrowed spectrum.
In Section \ref{sec:sbc}, we present our results for a simulation-based calibration study \cite{Talts:2020, McLeod:2021}.
In Section \ref{sec:priors}, prior distributions and other modelling choices are detailed.
In Section \ref{sec:results}, we present experimental results for our synthetic and real spectroscopic data sets.
Finally, in Section \ref{sec:conclusions}, we discuss our conclusions and consider future directions for research.
\section{Statistical line narrowing model}
\label{sec:model}
We observe $K$ spectral measurements consisting of $N$ line shapes with additive errors
\begin{equation}
    y_k = y(\nu_k) = f\left( \nu_k; \delta_N(\nu_k; \boldsymbol{a}, \boldsymbol{l}), \bm\theta\right) + \epsilon_k,
    \label{eq:dataModel}
\end{equation}
where $y_k \in \mathbb{R}$ denotes a discretized measurement at a wavenumber location $\nu_k \in \{ \nu_1, \dots, \nu_K \}$ in the space of wavenumbers $\mathcal{S} \subset \mathbb{R}_+$.
Wavenumbers are measured in inverse centimetres (cm$^{-1}$), while the units of the intensities are application-dependent.
In this paper, $y_k$ is measured in scientific arbitrary units (a.u.). The space $\mathcal{S}$ has been discretized with sampling resolution $ h > 0 $ where $ h = \vert \nu_{k+1} - \nu_k \vert$.
We denote the vector of all measurements as $\boldsymbol{y} := (y_1, \dots, y_K)^T$.
The continuous spectral density model $f(\cdot)$ and its parameters $\bm{a}, \bm{l}, \bm\theta$ are described in detail below.
The Gaussian measurement error is $\epsilon_k \sim \mathcal{N}(0, \sigma_\epsilon^2)$ with variance assumed known and zero mean.

The spectral density model is given by
\begin{equation}
    f( \nu; \delta_N(\nu; \boldsymbol{a}, \boldsymbol{l}), \bm\theta) = \sum_{n=1}^N a_n \mathcal{K}_\bullet(\nu - l_n; \bm\theta)= \mathcal{K}_\bullet(\nu; \bm\theta) \ast \delta_N(\nu; \boldsymbol{a}, \boldsymbol{l}),
    \label{eq:spectrumModel}
\end{equation}
where $\ast$ denotes convolution with respect to $\nu$ and  
\begin{equation}
    \delta_N(\nu; \boldsymbol{a}, \boldsymbol{l}) := \sum\limits_{n = 1}^N a_n \delta(\nu - l_n),
\end{equation}
which is a linear combination of $N$ Dirac delta functions $\delta(\nu)$ at locations $\boldsymbol{l} = (l_1, \dots, l_N)^T$, with amplitudes $\boldsymbol{a} = (a_1, \dots, a_N)^T$.

\begin{figure}
    \centering
    \includegraphics[width = \textwidth]{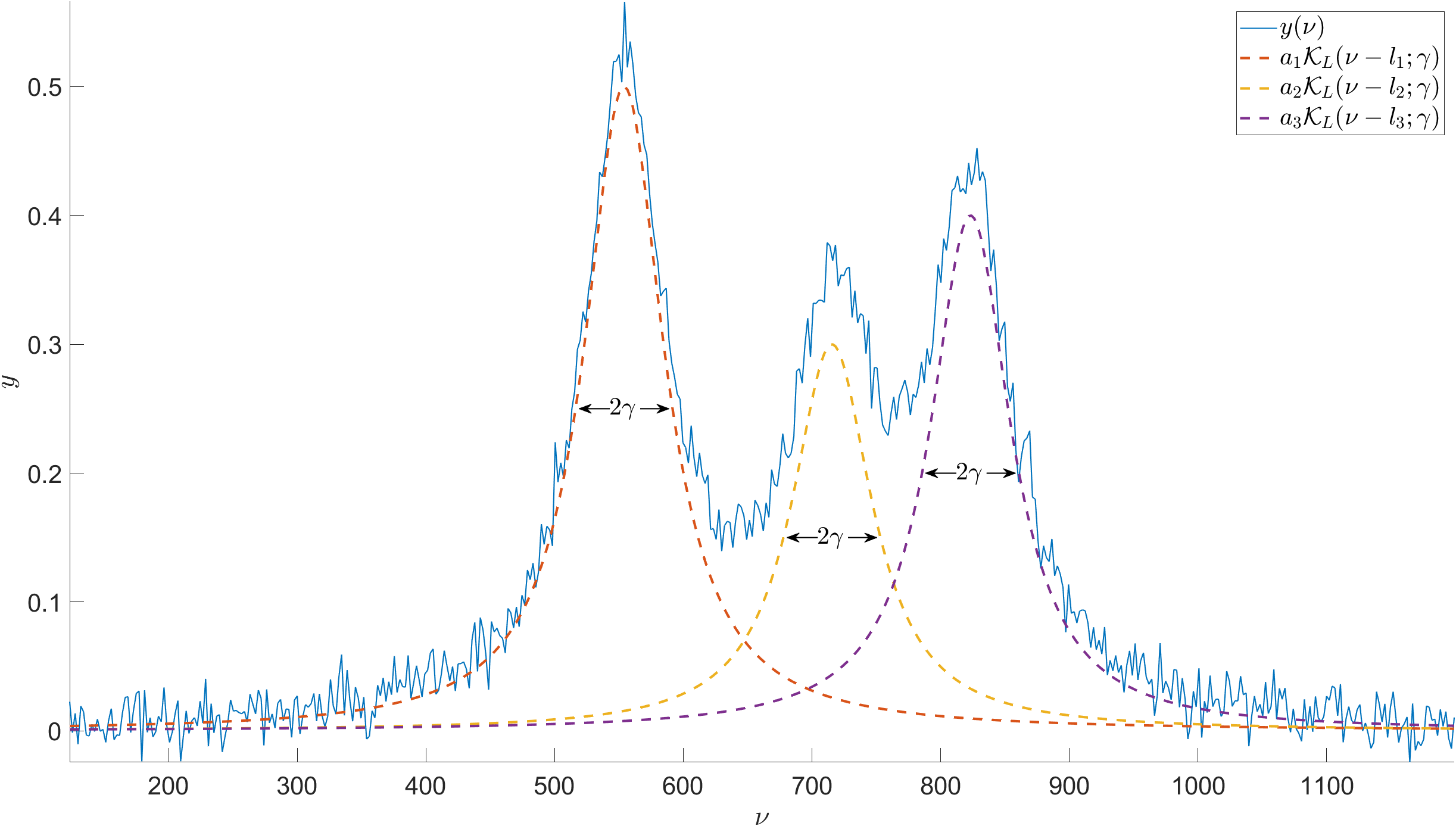}
    \includegraphics[width = \textwidth]{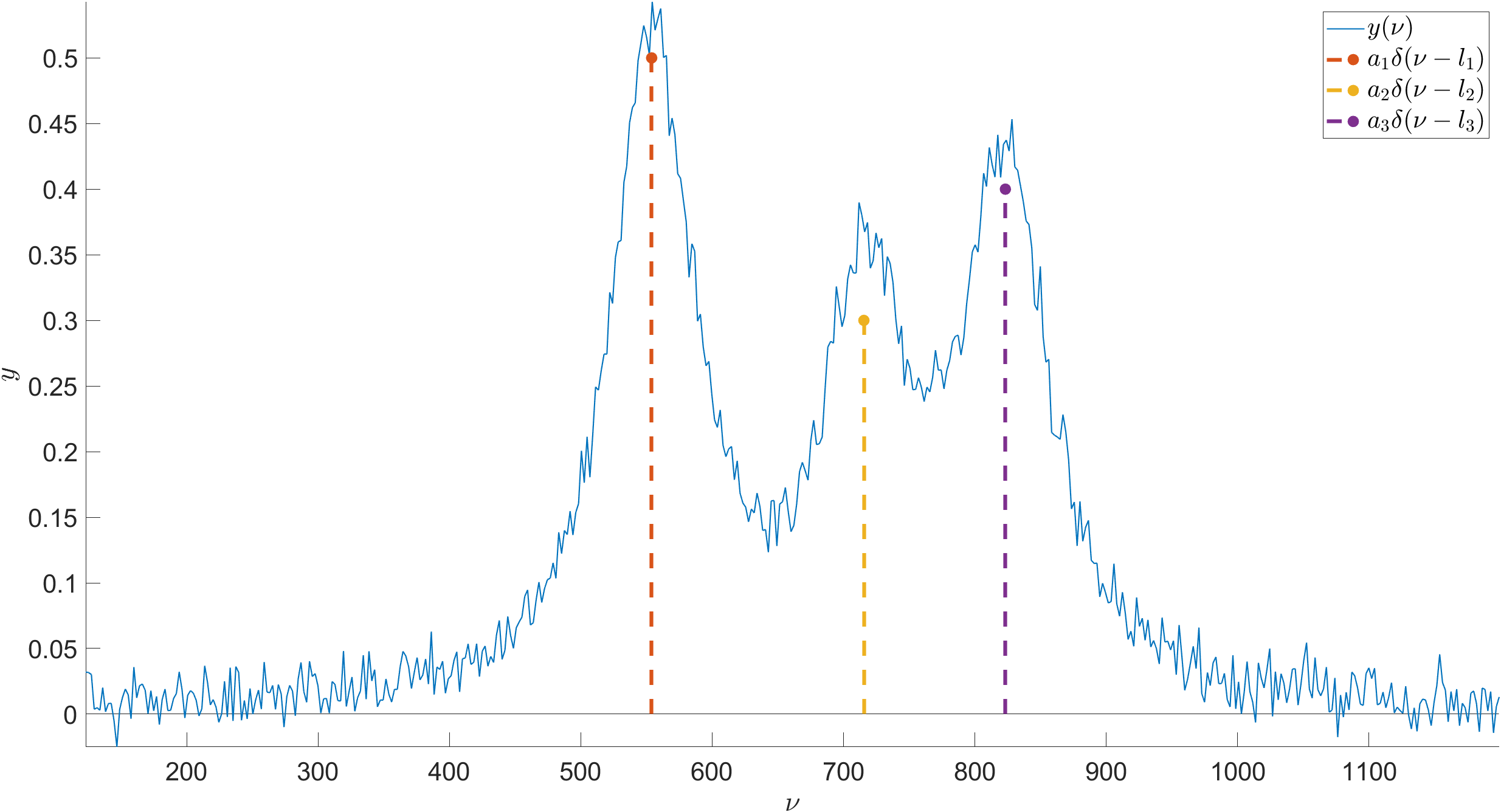}
    \caption{On top, a spectrum (blue) consisting of $N = 3$ Lorentzian line shapes located at locations $( l_1, l_2, l_3)^T$ shown in red, yellow, and purple, respectively. Upon successful line narrowing, or deconvolution, we would obtain three individual Dirac delta functions located at $( l_1, l_2, l_3)^T$. The aim of this paper is to construct approximate samples for the Dirac delta functions using linear prediction which are further modelled as a log-Gaussian Cox process.}
    \label{im:exampleLineshapes}
\end{figure}

The common line shape, or kernel, $ \mathcal{K}_\bullet(\nu; \bm\theta)$ is parameterized according to a vector of $P$ line shape parameters, $\bm\theta = (\theta_1, \dots, \theta_P)^T$.
We consider two common line shapes relevant for spectroscopic applications, the Lorentz line shape
\begin{equation}
    \mathcal{K}_L(\nu; \gamma) =  \frac{1}{\pi \gamma}\frac{\gamma^2}{ \nu^2 + \gamma^2},
    \label{eq:lorentz}
\end{equation}
with $P=1$ scale parameter $\theta_1 \equiv \gamma$,
and the Voigt line shape
\begin{equation}
\begin{split}
    \mathcal{K}_V( \nu; \sigma, \gamma) & = \mathcal{K}_L(\nu; \gamma) \ast \mathcal{K}_G(\nu; \sigma)
    = \frac{1}{\pi \gamma}\frac{\gamma^2}{ \nu^2 + \gamma^2} * \frac{1}{\sqrt{2\pi \sigma^2}} \exp\left( - \frac{ \nu^2}{2\sigma^2} \right),
\end{split}
\label{eq:voigt}
\end{equation}
with $P=2$ parameters, $\theta_1 \equiv \gamma$ and $\theta_2 \equiv \sigma$, which denote the scale parameters of Lorentzian and Gaussian line shapes, respectively.
In Raman spectroscopy, the Lorentzian line shape results from collisional broadening between molecules, while the Gaussian line shape results from Doppler broadening.
When both of these mechanisms are active simultaneously, this results in a Voigt line shape \cite{Diem:2015}.
Illustrations of Lorentzian line shapes and the corresponding Dirac delta functions are shown in Figure \ref{im:exampleLineshapes}.

The preceding construction gives a generative model for data $y_k$, where the parameters are known.
In order to learn the parameters from observed data, a different but related perspective is needed.
In the Fourier domain, the Dirac delta functions can be represented in terms of the Fourier self-deconvolution signal \cite{fsdarticle}
\begin{equation}
\begin{split}
     \xi( \omega; \bm{a}, \bm{l}, \bm\theta) := \mathcal{F} \left\{\sum\limits_{n=1}^N a_n\delta(\nu - l_n) + a_n\delta(\nu + l_n)\right\} &= 2\sum\limits_{n=1}^N a_n \cos( 2\pi \omega l_n) \\ &\approx \frac{ \mathcal{F} \left\{y(\nu) + y(-\nu)\right\} }{ \mathcal{F}\left\{ \mathcal{K}_\bullet(\nu; \bm\theta) \right\} },
\end{split}
\label{eq:FSD}
\end{equation}
where $\mathcal{F}$ is the Fourier transform and $ \omega $ denotes the Fourier-transformed variable.
Convolution in $\nu \in \mathcal{S}$ corresponds to multiplication in the Fourier domain, $\omega \in \Omega$ \cite{kauppAsump}.
Therefore, when we divide by $\mathcal{F}\left\{ \mathcal{K}_\bullet(\nu; \bm\theta) \right\}$ in Equation~\eqref{eq:FSD}, this operation corresponds to deconvolution.
The continuous Fourier transform of the Dirac delta function is equal to the cosine function.
In practice, the domain $\mathcal{S}$ has been discretized, so we employ the discrete Fourier transform (DFT).
We first reflect the observation vector $\bm{y}$ about the origin to obtain $y(+\nu_k)$ and $y(- \nu_k)$ for $k=1,\dots,K$ to produce an even function.
This is so that the DFT results in a real-valued Fourier representation.
However, the presence of noise in $\bm{y}$ means that the deconvolution is only approximate.
This approximation error balloons as the Fourier transform of the kernel $\mathcal{K}_\bullet(\nu; \bm\theta)$ in the denominator of Equation~\eqref{eq:FSD} approaches zero.

The aforementioned approximations result in only part of the discrete Fourier self-deconvolution signal being useful.
This useful part is truncated from the full-length signal to length $2 \leq M \leq K$ first samples.
The $M$ samples are then used to linearly predict the signal to the original length, $K$.
The linear prediction is performed as a recursive one-point extrapolation
\begin{equation}
    \xi_{\rm LP}( \omega_k; \bm{\theta}, M) = \Delta\omega \sum\limits_{i = 1}^M r_i\; \xi( \omega_{k - i}; \bm{a}, \bm{l}, \bm{\theta} ),
    \label{eq:linearPrediction}
\end{equation}
for $ k = M + 1, \dots, K$, where $\Delta\omega$ is the equidistant spacing and $r_i$ is the $i$th LOMEP impulse response coefficient. 
We use Levinson-Durbin recursion and Burg's formula to estimate the impulse response $\bm{r} = ( r_1, \dots, r_M)^T$ \cite{Kauppinen:91,kauppAsump}.

Under the assumptions given in Equation \eqref{eq:dataModel} the exact likelihood of the observed spectrum is
\begin{equation}
     \mathcal{L}( \boldsymbol{y} \mid \bm{f}( \bm\nu; \bm{a}, \bm{l}, \bm\theta ), \sigma_\epsilon^2) =  \prod\limits_{k = 1}^K \mathcal{N} \left( y_k; f\left( \nu_k; \delta_N(\nu_k; \boldsymbol{a}, \bm{l}), \bm\theta\right),  \sigma_\epsilon^2 \right),
     \label{eq:exactLikelihood}
\end{equation}
where $\bm{f}(\bm\nu; \bm a, \bm{l}, \bm\theta) = ( f\left( \nu_1; \delta_N(\nu_1; \boldsymbol{a}, \bm{l}), \bm\theta\right), \dots, f\left( \nu_K; \delta_N(\nu_K; \boldsymbol{a}, \bm{l}), \bm\theta\right) )^T$ is the underlying spectral signal evaluated at wavenumbers $\bm\nu = (\nu_1, \dots, \nu_K)^T$.
However, in practice this signal is unknown and cannot be directly observed. Instead, we approximate it using
\begin{equation}\label{eq:approxInverse}
    f\left( \nu_k; \delta_N(\nu_k; \boldsymbol{a}, \bm{l}), \bm\theta\right) \approx g(\nu_k, \bm\theta, M) := \mathcal{F}^{-1}\Bigl\{ \xi_{\text{LP}}( \bm{\omega}; \bm\theta, M) \mathcal{F}\left\{ \mathcal{K}_\bullet(\bm\nu, \bm\theta) \right\}\Bigl\}(\nu_k),
\end{equation}
where $\mathcal{F}^{-1}$ denotes the inverse DFT.

\begin{figure}
    \centering
    \includegraphics[width = \textwidth]{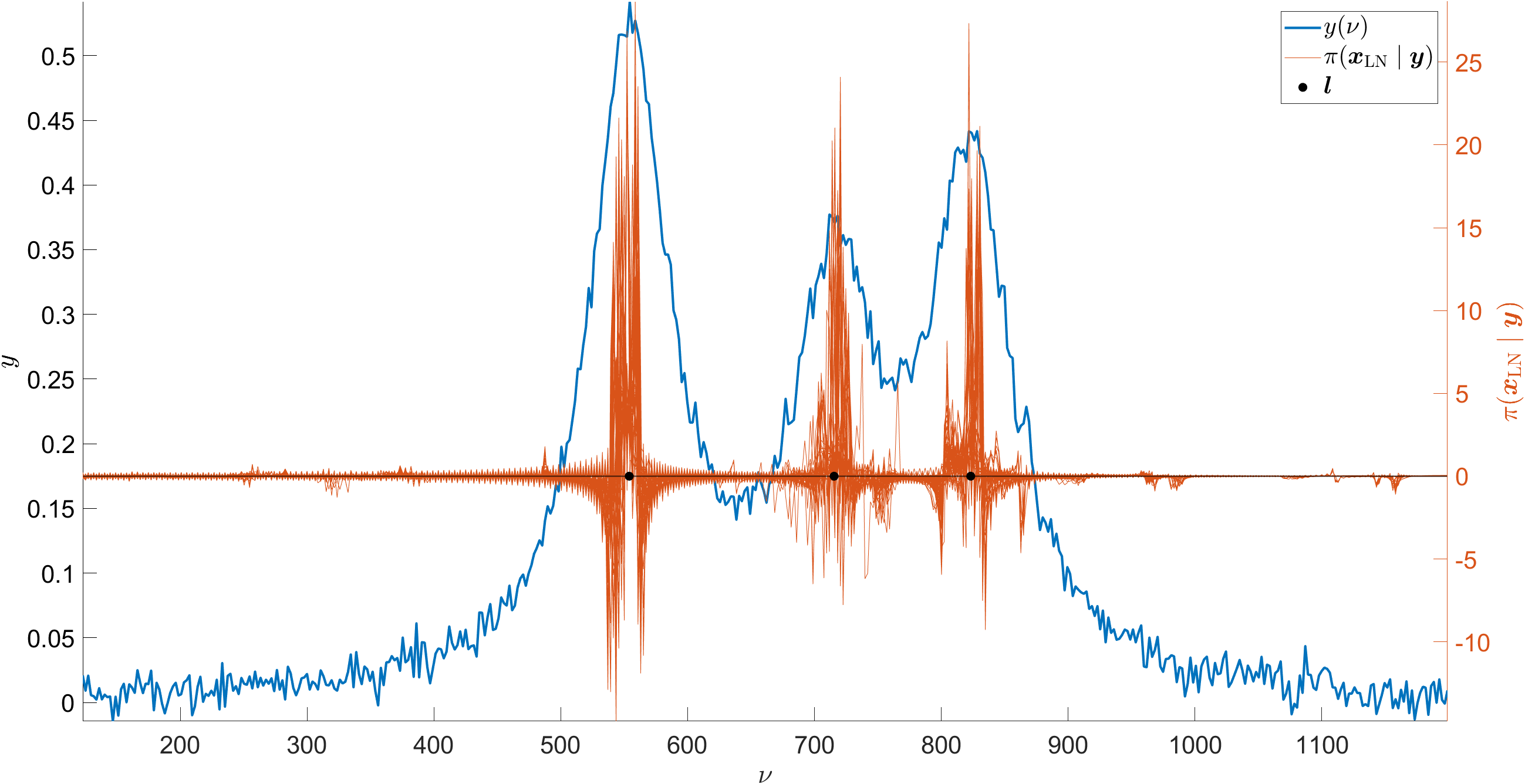}
    \includegraphics[width = \textwidth]{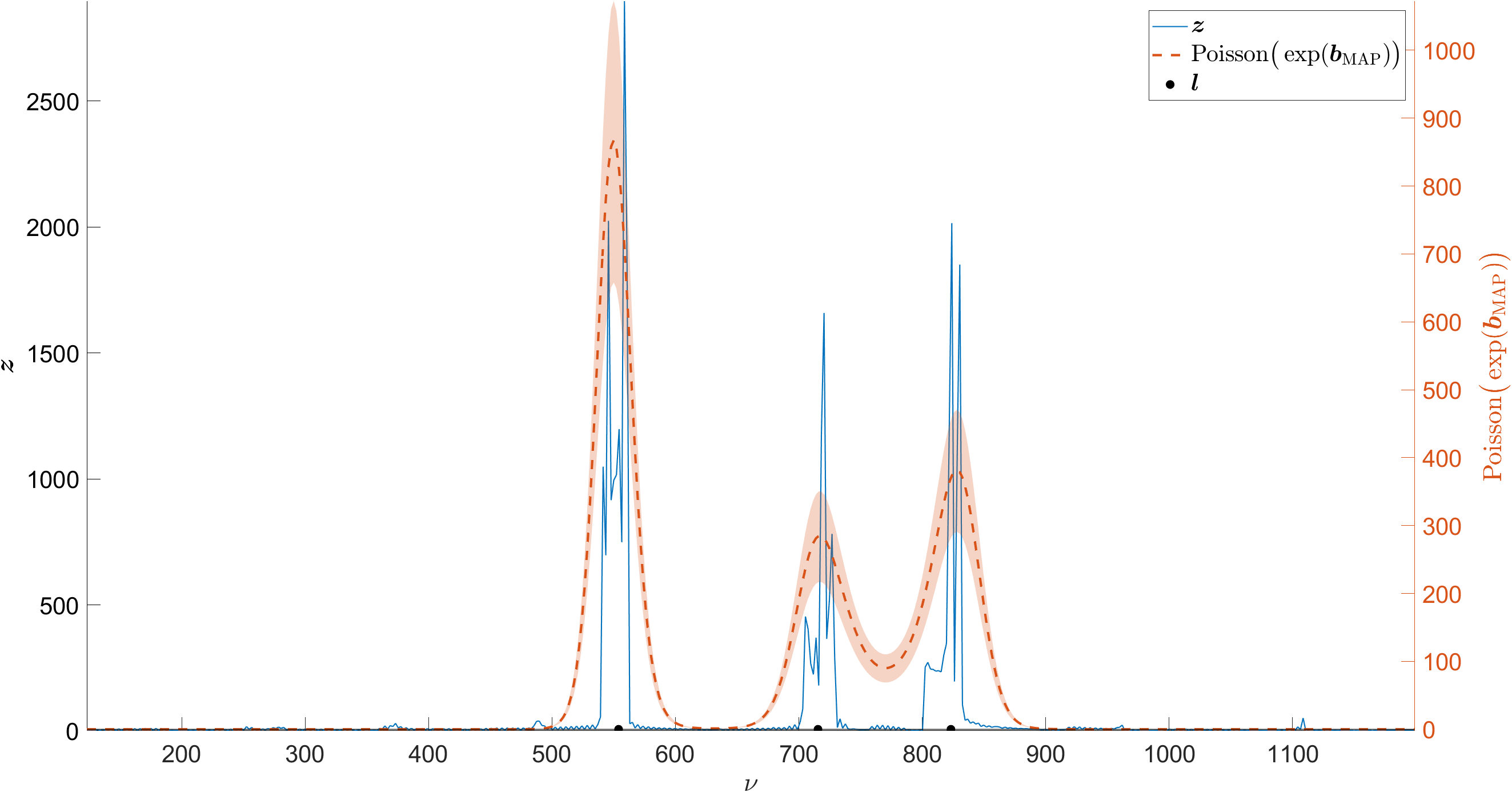}
    \includegraphics[width = \textwidth]{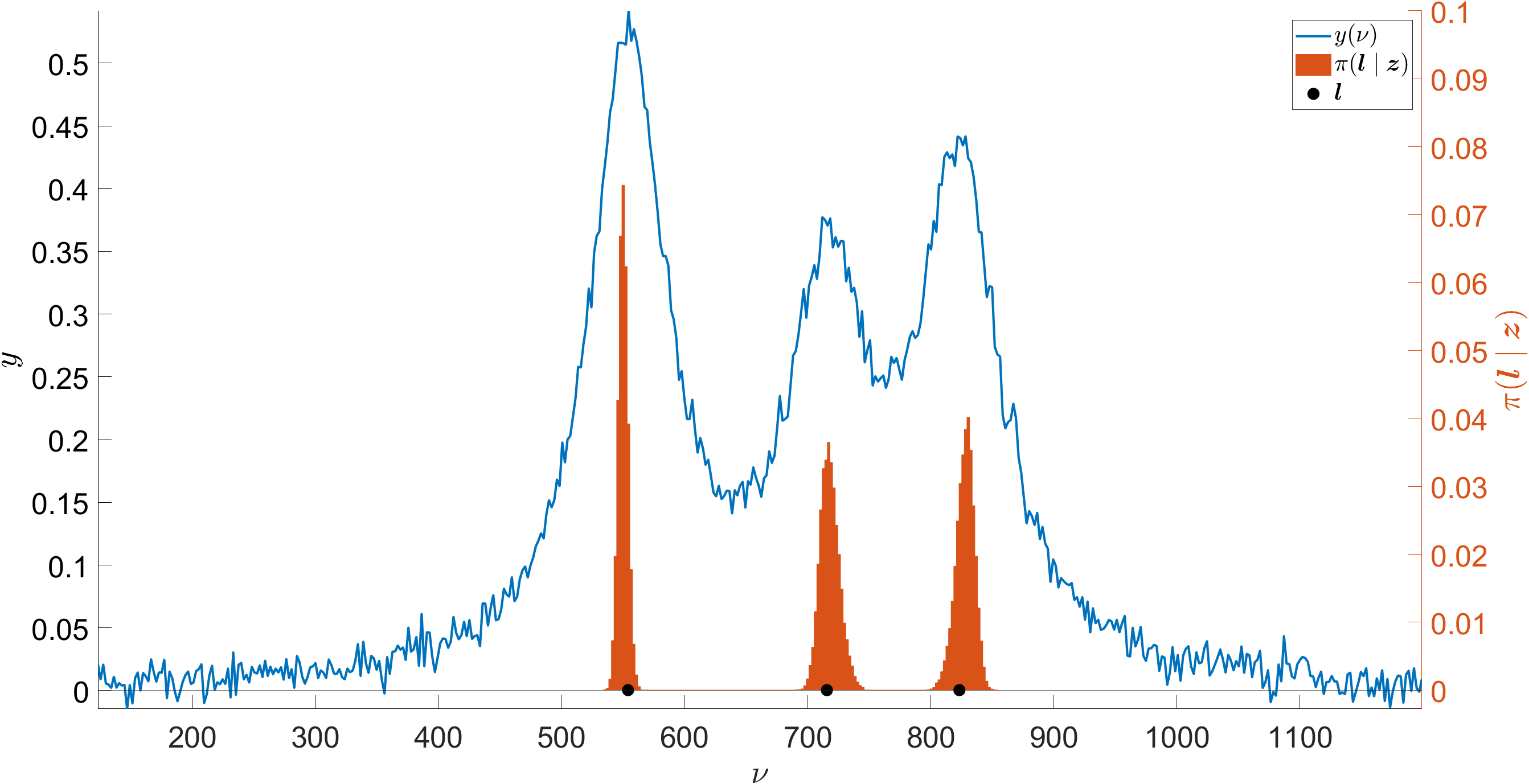}
    \caption{On top, a summary of the distribution of posterior samples from $\pi( \bm{x}_{\text{LN}} \mid \boldsymbol{y})$ for the spectrum in Figure \ref{im:exampleLineshapes}. In the middle, marginalized posterior samples according to Eq.~\eqref{eq:marginalizedSamplesLOMEP} and the corresponding LGCP estimate. At the bottom, the posterior $\pi( \bm{l} \mid \bm{z})$ for the line shape locations $\bm{l}$ obtained by sampling the LGCP local maxima.}
    \label{im:exampleResultLGCP}
\end{figure}

For brevity, henceforth we use a shorthand notation for the line-narrowed spectrum $\bm{x}_{\text{LN}} := \left( x_{\text{LN}}( \nu_1; \bm\theta, M), \dots, x_{\text{LN}}( \nu_K; \bm\theta, M) \right)^T = \mathcal{F}^{-1}\left\{ \bm\xi_{\text{LP}}( \bm{\omega}; \bm\theta, M) \right\}$.
The likelihood \eqref{eq:exactLikelihood} can then be approximated by
\begin{equation}
    \mathcal{L}\left( \boldsymbol{y} \mid \bm{f}( \bm\nu; \bm{a}, \bm{l}, \bm\theta), \sigma_\epsilon^2 \right) \approx \widetilde{\mathcal{L}}\left( \boldsymbol{y} \mid \bm{x}_{\text{LN}}, \sigma_\epsilon^2 \right) = \prod\limits_{k = 1}^K \mathcal{N}\left( y_k; g(\nu_k, \bm\theta, M), \sigma_\epsilon^2 \right).
\label{eq:quasiLike}
\end{equation}
The idea of approximating the likelihood using a Fourier transform is similar in spirit to the Whittle quasi-likelihood \cite{Whittle:1953}, but in our case $\widetilde{\mathcal{L}}$ also involves  deconvolution, truncation, and linear prediction.
An example of $\bm{x}_{\text{LN}}$ is shown at the top of Figure~\ref{im:exampleResultLGCP}, corresponding to the spectrum in Figure~\ref{im:exampleLineshapes}.
The problems of peak splitting, ringing artifacts and negative values common to LOMEP are clearly evident in this figure.

Given our quasi-likelihood, the approximate posterior distribution can be formulated as
\begin{equation}
    \pi\left( \bm{x}_{\text{LN}}, \bm\theta, M \mid \boldsymbol{y}, \sigma_\epsilon^2 \right) \propto \widetilde{\mathcal{L}}\left( \boldsymbol{y} \mid \bm{x}_{\text{LN}}, \sigma_\epsilon^2\right) \,\pi_0(\bm{x}_{\text{LN}} \mid \bm\theta, M) \pi_0( \bm\theta ) \pi_0( M ) ,
    \label{eq:posterior}
\end{equation}
where $\pi_0( \bm\theta )$, $\pi_0( M )$, and $\pi_0(\bm{x}_{\text{LN}} \mid \bm\theta, M)$ denote prior distributions for the line width, the Fourier self-deconvolution cut-off point, and the line-narrowed spectrum, respectively.
The probability distribution defined in Equation \eqref{eq:posterior} is not available in closed form and thus we require Bayesian computational methods to obtain stochastic samples from this distribution.
The SMC algorithm that we use for this purpose is similar to the algorithm in \cite{Harkonen:2020} but with different likelihood and prior distribution formulations.
This is similar in spirit to methods such as SMC-ABC \cite{DelMoral:2012}, which also targets an approximate posterior.
The quasi-likelihood approximation \eqref{eq:quasiLike} that we have introduced above substantially reduces the computational cost, as we discuss in the following section.

\section{Sequential Monte Carlo}
\label{sec:smc}
SMC, also known as the particle filter, or sequential importance sampling with resampling, is a class of algorithms for Bayesian computation that are very widely used for statistical signal processing and time series analysis.
For a general overview of SMC methods, we recommend \cite{Chopin:2020,Sarkka:2013}.
The SMC algorithm that we introduce here employs sequential importance sampling from a series of tempered probability distributions $\pi^{(0)}, \pi^{(1)}, \dots, \pi^{(T)}$ to ultimately obtain samples from the desired posterior distribution defined in Equation \eqref{eq:posterior}.

We construct the distribution at step $t$ of the tempering sequence as
\begin{equation}
    \pi^{(t)}\left( \bm{x}_{\text{LN}}, \bm\theta, M \mid \boldsymbol{y}, \sigma_\epsilon^2 \right) \propto \widetilde{\mathcal{L}}\left( \boldsymbol{y} \mid \bm{x}_{\text{LN}}, \sigma_\epsilon^2 \right)^{\kappa(t)} \,\pi_0(\bm{x}_{\text{LN}} \mid \bm\theta, M) \pi_0( \bm\theta ) \pi_0( M )
    \label{eq:temperedPosterior}
\end{equation}
where the superscript $(t)$ denotes the iteration of the SMC algorithm and the tempering parameter $\kappa^{(t)}$ is chosen such that $\kappa^{(t - 1)} < \kappa^{(t)} < \kappa^{(t + 1)} < \dots \leq 1$ with $\kappa^{(0)} = 0$.
In Equation \eqref{eq:temperedPosterior}, the initial state of the particle distribution is equal to the joint prior for $\bm\theta$, $M$, and $\bm{x}_{\text{LN}}$: that is $\pi^{(0)} = \pi_0(\bm{x}_{\text{LN}} \mid \bm\theta, M) \pi_0( \bm\theta ) \pi_0( M )$.
At each subsequent iteration, the particles are updated so that the intermediate tempering distribution approaches the targeted posterior  given in Equation \eqref{eq:posterior}.

We determine the tempering schedule adaptively, as in \cite{Harkonen:2020,Moores:2018}, such that the relative decrease in effective sample size (ESS), defined as
\begin{equation}
    J_{\text{ESS}}^{(t)} = \frac{1}{\sum\limits_{j = 1}^J\left( w_j^{(t)} \right)^2},
    \label{eq:ess}
\end{equation}
is approximately some predefined learning rate $\eta$ between SMC iterations. Here, $w_j^{(t)}$ is the importance sampling weight for the $j$th particle $( \bm{x}_{\text{LN}, j}, \bm\theta_j, M_j)$, where $j = 1,\dots,J$. The incremental unnormalized weights in the SMC algorithm are
\begin{equation}
    W_j^{(t)} \propto \frac{ \widetilde{\mathcal{L}}\left( \boldsymbol{y} \mid \bm{x}_{\text{LN}}, \sigma_\epsilon^2 \right)^{\kappa(t)} }{ \widetilde{\mathcal{L}}\left( \boldsymbol{y} \mid \bm{x}_{\text{LN}}, \sigma_\epsilon^2\right)^{\kappa(t - 1)} } w_j^{(t - 1)},
    \label{eq:weight_update}
\end{equation}
with the normalized weights given by $w_j^{(t)} = W_j^{(t)} / \sum_{j=1}^J W_j^{(t)}$.

As the number of iterations increases, the weights will gradually become concentrated on a small number of particles, hence $J_{\text{ESS}}^{(t)}$ decreases. 
This can eventually result in a single particle dominating the inference.
To avoid this tendency, we employ two additional steps: resampling and mutation of the particles.
Resampling is initiated when the ESS drops below a set threshold $J_{\text{min}}$.
Each particle is resampled with replacement, using probabilities equal to $w_j^{(t)}$.
This often means that particles with the most weight are sampled multiple times, creating duplicates.
After resampling, the weights are all reset to $w_j^{(t)} = \frac{1}{J}$.
Markov chain Monte Carlo (MCMC) is then used to update the particles to move duplicate particles to different states.
The target distribution for the MCMC is defined by the tempered posterior distribution at iteration $t$ as given in Equation \eqref{eq:temperedPosterior}.
We present pseudo-code for our SMC sampled line narrowing method in Algorithm \ref{alg:smc_lna}.
\begin{algorithm}
\caption{Sequential Monte Carlo sampled line narrowing.}
\label{alg:smc_lna}
\begin{algorithmic}
\State \textbf{Initialize:}
	\State \hspace{\algorithmicindent} Set $t = 0$ and $\kappa^{(t)} = 0$.
	\State \hspace{\algorithmicindent} Sample $J$ particles independently from the prior $ \pi_0( \bm\theta ) \pi_0( M )$.
	\State \hspace{\algorithmicindent} Compute a line-narrowed spectrum $ \bm{x}_{\text{LN}, j}$ for each particle $(\bm\theta_j, M_j)$.
	\State \hspace{\algorithmicindent} Compute the quasi-likelihood $\widetilde{\mathcal{L}}( \boldsymbol{y} \mid \bm{x}_{\text{LN}, j}, \sigma_\epsilon^2)$ for each particle.
    \State \hspace{\algorithmicindent} Set particle weights $w_j^{(t)} = \frac{1}{J}$.
\While{$\kappa^{(t)} < 1$}
    \State $t = t + 1$.
    \State Determine $\kappa^{(t)}$ according to the learning rate $\eta$.
    \State Update particle weights $w_j^{(t)}$ according to Equation \eqref{eq:weight_update}.
    \State Compute the effective sample size $J_{\text{ESS}}^{(t)}$ using Equation \eqref{eq:ess}.
    \If{$J_{\text{ESS}}^{(t)} < J_{\text{min}}$}
    	\State Resample particles according to their weights.
    	\State Set particle weights $w_j^{(t)} = \frac{1}{J}$.
    \EndIf
    \State Update particles with MCMC targeting the tempered posterior given by \eqref{eq:temperedPosterior}.
\EndWhile
\end{algorithmic}
\end{algorithm}
After the final iteration is complete and $\kappa^{(T)} = 1$, the distribution of the $J$ particles represent samples from the target posterior distribution, Equation \eqref{eq:posterior}.

Linear prediction requires Fast Fourier transforms (FFT) of size $K$, which are $\mathcal{O}(K \log( K ))$, solving a Toeplitz matrix at worst of size $M_{\rm max} \times M_{\rm max}$ where $M_{\rm max}$ is the upper bound defined by the prior $\pi_0(M)$ resulting in $\mathcal{O}( M_{\rm max}^2 )$, and the complexity of the prediction is at worst $K - M$ values with $M$ impulse response coefficients which is $\mathcal{O}( ( K - M )M )$.
For large enough $K$ and $M_{\rm max} \ll K $, the cost of linear prediction is dominated by the complexity of the FFT, $\mathcal{O}( K \log(K) )$.
The linear prediction is performed at each step $t$ of the SMC sampling $N_{\rm MCMC}$ times where $N_{\rm MCMC}$ is the number of MCMC iterations.
This leads to a total of $TN_{\rm MCMC}$ likelihood evaluations, resulting in a total complexity of $\mathcal{O}( KT \log(K))$.

However, the marginal posterior $\pi( \bm{x}_{\text{LN}} \mid \boldsymbol{y})$ for the line-narrowed spectrum is lacking in physical interpretability.
This is due to the phenomenon of ``peak splitting,'' where individual peaks in the true spectrum $\bm{f}(\bm\nu; \bm\theta)$ are split into two or more peaks in the line-narrowed spectrum $\bm{x}_{\text{LN}}$ \cite{Kauppinen:1992}.
The line-narrowed spectra are approximations of Dirac delta functions, so their marginalization  is difficult to visualize accurately.
We show a simplified illustration of the marginalized posterior for  $\bm{x}_{\text{LN}}$ in Figure~\ref{im:exampleResultLGCP}.
This leads us to consider ``smoothing'' the peaks, or more accurately, estimating the underlying distribution of peak locations $l_n$, which we assume to be distributed according to a log-Gaussian Cox process.
\section{Log-Gaussian Cox Process}
\label{sec:lgcp}
A point process is a countable collection of random locations $\{ l_1, l_2, \dots \}$ within some space, $\mathcal{S}$.
In our case, we take $\mathcal{S}$ to be the one-dimensional space of wavenumbers $\mathbb{R}_+$, or more specifically the continuous interval bounding the fingerprint region for organic molecules, $l_n \in [120, 1200]$.
Closely related to this random set of points is the counting process $\varphi(A)$ for measurable subsets $A \subseteq \mathcal{S}$.
For example, if $A = [800, 900]$ and there are two peaks with locations $l_1 = 810$ and $l_2 = 850$, then $\varphi(A)$ = 2.

A Poisson process is a type of point process that satisfies the following properties \cite{Last:2017}:
\begin{enumerate}
    \item[(i)] Whenever $A_1, A_2, \dots, A_Q \subset \mathcal{S}$ are disjoint, then $\varphi(A_1), \varphi(A_2), \dots, \varphi(A_Q)$ are independent random variables. That is,
\begin{equation}\label{eq:PoisIndep}
\mathbb{P}\left(\bigcap_{i \in \mathcal{I}} \{\varphi(A_i) = z_i\}\right) = \prod_{i \in \mathcal{I}} \mathbb{P}(\varphi(A_i) = z_i) , \quad z_i \in \mathbb{N} \cup \{0\},\, \mathcal{I} \subseteq \{1,\dots,Q\} .
\end{equation}
    \item[(ii)] The random variable $\varphi(A)$ follows a Poisson distribution, with expectation $$\mathbb{E}[\varphi(A)] = \Lambda(A) ,$$
\end{enumerate}
where $\Lambda(A)$ is known as the intensity measure.
Let $\lambda$ be the Radon-Nikod{\'y}m derivative of $\Lambda$, so that $\Lambda(A) = \int_A \lambda(a) \,\mathrm{d}a$, then $\lambda$ is known as the intensity function.

A log-Gaussian Cox process (LGCP) is a {\em doubly}-stochastic point process, where the intensity function $\lambda$ of an inhomogeneous Poisson process is modelled as a stochastic process, or a random function, in itself.
Specifically, the logarithm of the intensity is considered to follow a Gaussian process (GP) \cite{Moller:1998},
\begin{equation}
\log \lambda( \nu ) \sim \text{GP}\left( \mathbf{0},\; \Sigma( \nu, \nu'; \bm\psi) \right) ,
\end{equation}
where $\Sigma( \nu, \nu'; \bm\psi)$ is the covariance function of the GP with parameter vector $\bm\psi$, evaluated at locations $\nu, \nu' \in \mathcal{S}$.
We use a squared exponential covariance,
\begin{equation}
    \Sigma( \nu, \nu'; \bm\psi) = \sigma_\lambda^2 \exp\left( -\frac{1}{2}\frac{ \left( \nu - \nu' \right)^2 }{ \ell^2 } \right),
    \label{eq:sqExpGP}
\end{equation}
where $\sigma_\lambda$ is the standard deviation of the GP and $\ell$ is its length scale parameter, so that $\bm\psi = (\sigma_\lambda, \ell)^T$.

The exact likelihood of the LGCP is intractable for continuous $\mathcal{S}$, so we follow the advice of \cite{Moller:1998} and discretize the domain.
In our case, $\mathcal{S}$ is already partitioned into disjoint subsets $A_1, \dots, A_K$ at equally-spaced locations $\nu_1, \dots, \nu_K$, each being $h$ wavenumbers apart, so that $\bigsqcup_{k=1}^K A_k = \mathcal{S}$ and $| \mathcal{S} | = K h$.
This is a result of the measurement technology, a spectrometer is capable of measurement at a discretized set of measurement points.
This yields a natural, and unavoidable, discretization of the space $\mathcal{S}$ with spacing $h$.
In order to fit the LGCP, we need to translate the SMC samples for the line-narrowed spectrum $\bm{x}_{\text{LN},j}$ into approximate counts $z_k$ of the number of Dirac delta functions located inside each subset $A_k$.
However, these samples can have negative values due to the Gibbs phenomenon or ringing caused by the approximation of delta functions via Fourier transforms, as can be seen in Figure \ref{im:exampleResultLGCP}.
Thus, we discretize and translate the samples so that a LGCP can be utilized.
We initially marginalize the samples with respect to $\bm\theta$ and $M$ as
\begin{equation}
\bar{x}_k := \mathbb{E}_{\bm\theta, M}\left[    x_{\text{LN}}(\nu_k) \mid \boldsymbol{y} \right] \approx \frac{y_{\text{area}}}{J} \sum\limits_{j = 1}^J   \frac{  x_{\text{LN}, j}( \nu_k; \bm\theta_j, M_j) \mathds{1}_{x \geq 0}}{\sum_{k=1}^K x_{\text{LN}, j}( \nu_k; \bm\theta_j, M_j) \mathds{1}_{x \geq 0}} ,
\end{equation}
where $\bar{x}_k$ denotes the  posterior expectation for the line-narrowed spectrum at $\nu_k$, $y_{\text{area}} = \sum_{k = 1}^K y_k$ the area under the measurements, and $\mathds{1}_{x \geq 0}$ is an indicator function such that
\begin{equation}
 \mathds{1}_{x \geq 0} =
 \begin{cases}
    1, & x_{\text{LN}, j}( \nu_k; \bm\theta_j, M_j) \geq 0,\\
    0, & x_{\text{LN}, j}( \nu_k; \bm\theta_j, M_j) < 0.
 \end{cases}
\end{equation}
Next, we scale the values by a pre-defined constant of proportionality $C$ and round the values to the closest integers.
Given this, we obtain discretized and positive values $\bm{z} = ( z_1, \dots, z_K )^T$ constructed as
\begin{equation}
    z_k = \left\lfloor C \bar{x}_k  + \frac{1}{2} \right\rfloor
    \label{eq:marginalizedSamplesLOMEP}
\end{equation}
where $\lfloor \, x \, \rfloor$ denotes the greatest integer $\le x$.

Given the above, we are able to approximate the likelihood of the LGCP by
\begin{equation}
    \mathcal{L}(\bm{z} \mid \bm{b}) = \prod\limits_{k = 1}^K \frac{ \lambda(\nu_k)^{z_k}}{z_k !} \exp\left\{ \lambda(\nu_k) \right\},
    \label{eq:poisson_likelihood}
\end{equation}
where $\bm{b} = (\log \lambda(\nu_1), \dots, \log\lambda(\nu_K) )^T$.
In turn, the GP prior for $\bm{b}$ can be evaluated as
\begin{equation}
     \pi_0( \bm{b} \mid \bm\psi) = \frac{1}{\sqrt{(2\pi)^K}} \left\vert \Sigma( \bm\nu, \bm\nu; \bm\psi) \right\vert^{-1/2}
     \exp\left\{ 
     -\frac{1}{2} \bm{b}^T \Sigma( \bm\nu, \bm\nu; \bm\psi)^{-1} \bm{b} \right\},
    \label{eq:gp_loglikelihood}
\end{equation}
where $\left\vert \Sigma( \bm\nu, \bm\nu; \bm\psi) \right\vert$ denotes the determinant of the $K \times K$ covariance matrix.
The joint posterior distribution is then
\begin{equation}
    \pi( \bm{b}, \bm\psi \mid \bm{z} ) \propto \mathcal{L}(\bm{z} \mid \bm{b}) \,\pi_0( \bm{b} \mid \bm\psi) \,\pi_0( \bm\psi ),
    \label{eq:posteriorLGCP}
\end{equation}
where $\pi_0( \bm\psi )$ are priors for the parameters of the GP.
For inference of the posterior defined in \eqref{eq:posteriorLGCP}, we use MAP estimation via quasi-Newton optimization.
Specifically, the limited-memory  Broyden–Fletcher–Goldfarb–Shanno (L-BFGS) algorithm \cite{Liu:1989} as implemented in the GPstuff toolbox \cite{Vanhatalo:2013}.
The discretized and positive values $\bm{z}$ and corresponding MAP estimate for $\bm{b}$, along with 90\% posterior credible intervals, are illustrated in the middle in Figure~\ref{im:exampleResultLGCP}.

Finally, we construct a posterior distribution for the line shape locations $\bm{l}$ by sampling the estimated GP.
We use the local maxima of the sampled GP as our estimate for the peak locations $\bm{l}$.
That is, where
\begin{equation}
\frac{\partial \log\lambda}{\partial \nu} = 0 \text{ and } \frac{\partial^2 \log\lambda}{\partial \nu^2} < 0.
    \label{eq:GPderivatives}
\end{equation}
Note that realizations of the GP must be at least twice differentiable in order for \eqref{eq:GPderivatives} to be valid.
For example, the covariance function $\Sigma( \nu, \nu'; \bm\psi)$ could be a Mat{\'e}rn with smoothness parameter $\zeta = 5/2$.
Instead, we choose the squared exponential covariance \eqref{eq:sqExpGP}, which is equivalent to a Mat{\'e}rn in the limit as $\zeta \rightarrow \infty$, since this is guaranteed to produce smooth realizations.
We repeat the sampling for the GP local maxima $20000$ times to construct a posterior distribution $\pi( \bm{l} \mid \bm{z})$.
An example construction for this posterior is shown at the bottom of Figure \ref{im:exampleResultLGCP}.

\section{Simulation-based Calibration}
\label{sec:sbc}
We use simulation-based calibration (SBC) \cite{McLeod:2021, Talts:2020} to validate that our model is able to produce a consistent estimate of the number of peaks in the spectrum, $N$.
Under mild assumptions, Bayesian posterior distributions have the property of self-consistency.
This means that if we sample a parameter from its prior distribution,
\begin{equation}
    \theta^* \sim \pi_0(\theta),
\end{equation}
and then simulate data from the generative model
\begin{equation}
    \bm{y}^* \sim \mathcal{L}(\bm{y} \mid \theta^*),
\end{equation}
then we expect that the resulting  distribution $\pi(\theta \mid \bm{y}^*)$ of posterior probability should be concentrated in the vicinity of the true parameter value $\theta^*$.
This is particularly important in the context of our method, since we have replaced the true likelihood \eqref{eq:exactLikelihood} with an approximation based on LOMEP, $\widetilde{\mathcal{L}}\left( \boldsymbol{y} \mid \bm{x}_{\text{LN}}, \sigma_\epsilon^2 \right)$ \eqref{eq:quasiLike}.
We need to ensure that our approximation is accurate enough to still produce consistent estimates.

SBC involves generating multiple parameter values $\theta^*_1, \dots, \theta^*_S$ and corresponding synthetic datasets $\bm{y}^*_1, \dots, \bm{y}^*_S$ from their joint distribution,
\begin{equation}
    \left(\bm{y}^*_s, \theta^*_s\right) \sim \pi(\bm{y}, \theta),
\end{equation}
where $\pi(\bm{y}, \theta) = \mathcal{L}(\bm{y} \mid \theta)\ \pi_0(\theta)$. For our model, we first simulate a realization of a GP, $\bm{b}^*_s$.
We can then determine the peak locations $\bm{l}^*_s$ and number of peaks $N^*_s$ from the maxima of this function, as previously explained in Section~\ref{sec:lgcp}.
Next, we simulate amplitudes $\bm{a}^*_s$ and line shape parameter $\gamma^*_s$ for the peaks.
We can then evaluate $f\left( \nu;\ \delta_N(\nu; \bm{a}^*_s, \bm{l}^*_s),\ \bm\theta^*_s\right)$ at wavenumbers $\nu_1, \dots, \nu_K$ and add white noise to produce synthetic data $\bm{y}^*_s$ according to \eqref{eq:dataModel}.
In our case, we produce $S=100$ pairs of parameters and datasets from the joint distribution. The prior distributions for the parameters are described in Section~\ref{sec:priors}.

\begin{figure}
    \centering
    \hfill
    \includegraphics[width = 0.4\textwidth]{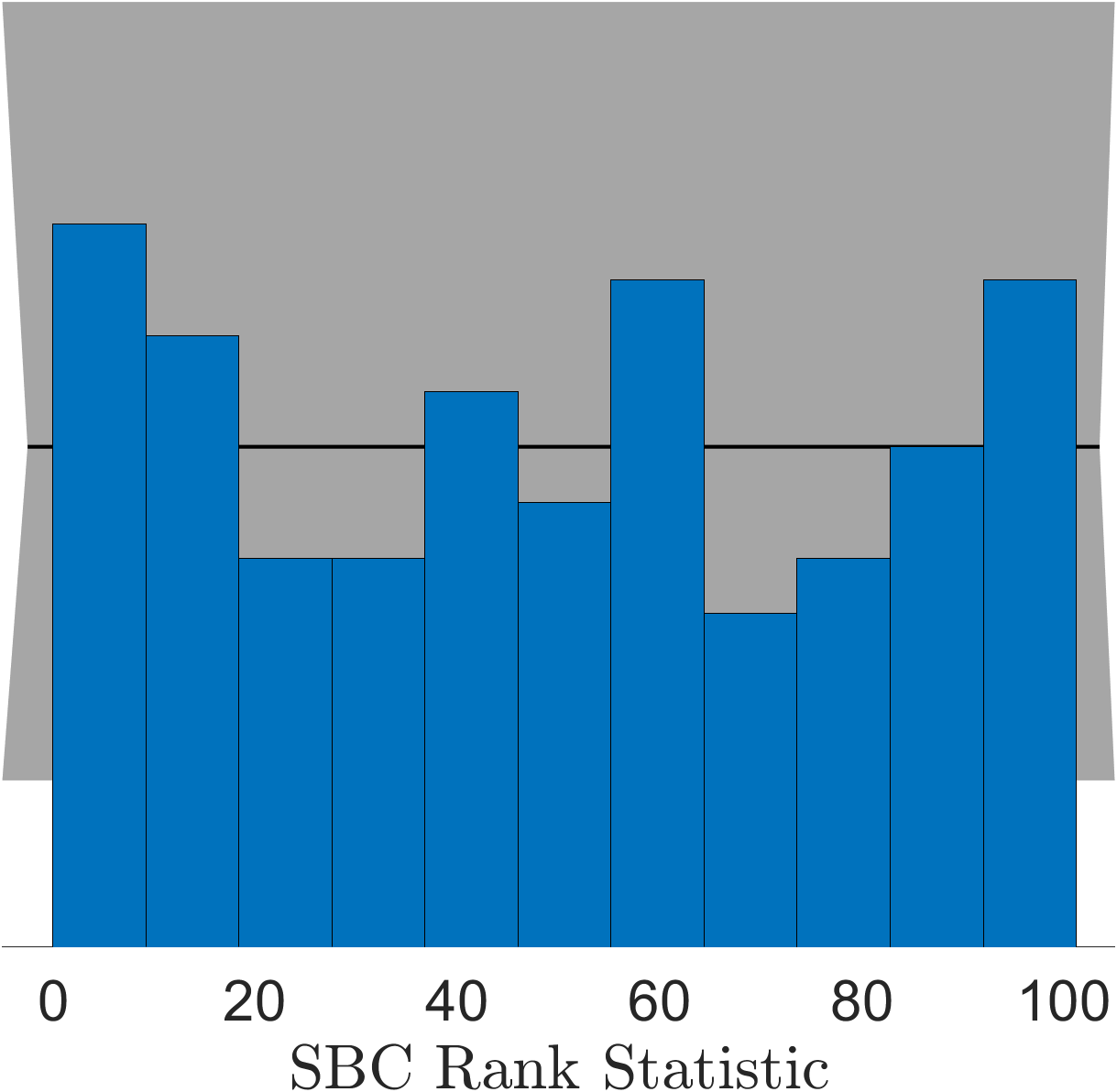}
    \hfill
    \hfill
    ~
    \caption{Simulation-based calibration histogram for the rank statistics of the true number of peaks $N^*_s$. The number of peaks were estimated as the number of local maxima in samples from the LGCP fits, using a GP length scale of $0.025$. The histogram shows a uniform distribution. The solid black line shows the expected value and the shaded gray areas show the 99\% confidence intervals.}
    \label{im:sbcBiasedRankStatistic}
\end{figure}
We run Algorithm~\ref{alg:smc_lna} to obtain samples from the approximate posterior for $\bm{x}_{\text{LN}}$ given $\bm{y}^*_s$, then use L-BFGS to fit the LGCP.
Finally, we repeatedly sample $\bm{b}_s^{(j)}$ from the GP for $j=1, \dots, J$ and find the maxima of each function to obtain posterior samples $\bm{l}_s^{(j)}$ and $N_s^{(j)}$.
We use $J$=20000 posterior samples as described in Section~\ref{sec:lgcp}.
Under Bayesian self-consistency, the rank $r_s$ of the true parameter $N^*_s$ should be uniformly-distributed under the corresponding posterior,
\begin{equation}
    \begin{split}
        r_s &= \text{rank}\left[ N^*_s ; ( N_s^{(1)}, \dots, N_s^{(J)}) \right]\\
&= \sum_{j=1}^J \mathds{1}\left[ N_s^{(j)} < N^*_s \right]
    \end{split}
\end{equation}
where $\mathds{1}[x]$ is the indicator function, therefore $r_s$ is a number between 0 and $J$. Figure~\ref{im:sbcBiasedRankStatistic} shows that these ranks $r_1, \dots, r_S$ are indeed uniformly-distributed, as required by SBC.

\section{Prior Distributions and Computational Details}
\label{sec:priors}
We use a continuous uniform distribution for the line width parameter $\gamma$ for both Lorentz and Voigt profiles, with $\sigma$ modelled as a truncated normal distribution conditional on $\gamma$.
A discrete uniform distribution is used for the cut-off parameter $M$.
We use a fixed value for the length scale parameter $\ell$ of the GP covariance function.
The prior for the GP covariance parameter $\sigma_\lambda$ is specified on a logarithmic scale.
We use Student's $t$ distributions $t( \mu, \delta^2, \nu_\text{Fr})$ parameterized according to mean, variance, and degrees of freedom.
We detail these prior distributions in Table \ref{tb:priors}.
\begin{table}
\caption{Prior distributions for the Lorentz and Voigt line shape parameters $\bm\theta$, the Fourier self-deconvolution cut-off parameter $M$, and the GP covariance parmeters $\psi$.}
\centering
\begin{tabular}{c c c}
\toprule
Prior & Lorentz & Voigt \\
\midrule
$\pi_0(\gamma)$ & $\mathcal{U}( 1, 30)$ & $\mathcal{U}( 1, 30)$ \\
$\pi_0(\sigma \mid \gamma)$ & NA & $\mathcal{N}_+( 0.5 \times \gamma, (0.05 \times \gamma)^2 )$ \\
$\pi_0(M)$ & $\mathcal{U}( 10, 80)$ & $\mathcal{U}( 10, 80)$ \\
$\pi_0(\log\{\sigma_\lambda\})$ & $t( 0, 100^2, 10)$ & $t( 0.01, 100^2, 10)$ \\
\bottomrule
\end{tabular}
\label{tb:priors}
\end{table}
More specifically we set $\pi_0(\gamma) = \mathcal{U}( 1, 30)$ and $\pi_0(M) = \mathcal{U}( 10, 80)$, meaning that the half-width at half-maximum (HWHM) of the peaks is limited to a range between 1 cm$^{-1}$ and 30 cm$^{-1}$ and the number of Fourier self-deconvolution points is limited to lay in the interval $[ 10, 80]$.
This prior on $M$ has an important effect in regularizing the discrete Fourier approximation given by Equation~\eqref{eq:approxInverse}.
If $M$ is too large, then the posterior for $\bm{x}_{LN}$ will be dominated by ringing artifacts and boundary effects, making it impossible to determine the peak locations.
The theoretical maximum number of peaks that can be represented by $g(\nu_k, \bm\theta, M)$ is $0.5 M$ \cite{kauppAsump}.
In practice, we recommend setting the upper bound for $\pi_0(M)$ to be at least four times the maximum number of peaks in the spectrum, but small enough that the effect of ringing artifacts is minimized.

The SMC was run with $J = 1000$ particles with residual resampling initiated when ESS falls below a threshold of $J_{\text{min}} = J / 2$.
We use a Metropolis-Hastings random walk kernel for the MCMC updates.
The MCMC proposals for new particles $(\gamma, M)_{1:J}^*$ are constructed as
\begin{equation}
\begin{split}
     (\bm\theta_j^*, \widetilde{M}_j^*) &= (\bm\theta_j, M_j) + \zeta,\\
     M_j^* &= \left\lfloor \widetilde{M} + \frac{1}{2} \right\rfloor + \zeta_M,
\end{split}
\end{equation}
where $\zeta \sim \mathcal{N}( 0, c\Sigma_{\bm\theta, N})$ with $\Sigma_{\bm\theta, N}$ denoting a empirical covariance of the current particles $(\bm\theta, N)_{1:J}$, scaled according to $c \in R_+$ such that the acceptance rate is approximately a pre-defined target acceptance rate, and with a discrete random walk $\zeta_M \sim \mathcal{U}( -1, 1)$.
New particles $ \bm{x}_{\text{LP},j}( \bm\nu, \bm\theta^*_j, M^*_j)$ are then be computed according to the linear prediction method.
The target acceptance rate was set to $0.30$ with 5 MCMC updates during each iteration step.
There exist methods of automatically determining the number of MCMC updates at each iteration step, see for example \cite{Dau:2022} and references therein.
\section{Results}
\label{sec:results}
In addition to the SBC study described in Section~\ref{sec:sbc}, we have also applied our SMC-sampled LOMEP algorithm to 4 other synthetic spectra, as well as a mineralogical Raman spectrum and two coherent anti-Stokes Raman spectra (CARS) of proteins.
All of the syntethic spectra were constructed using noise variance $\sigma_{\epsilon}^2 = 0.025^2$.
The results for the 4 synthetic spectra and the 2 proteins are available in the online supplementary material.

\begin{figure}
 {\hspace{-0.8cm} \includegraphics[width=0.94\textwidth]{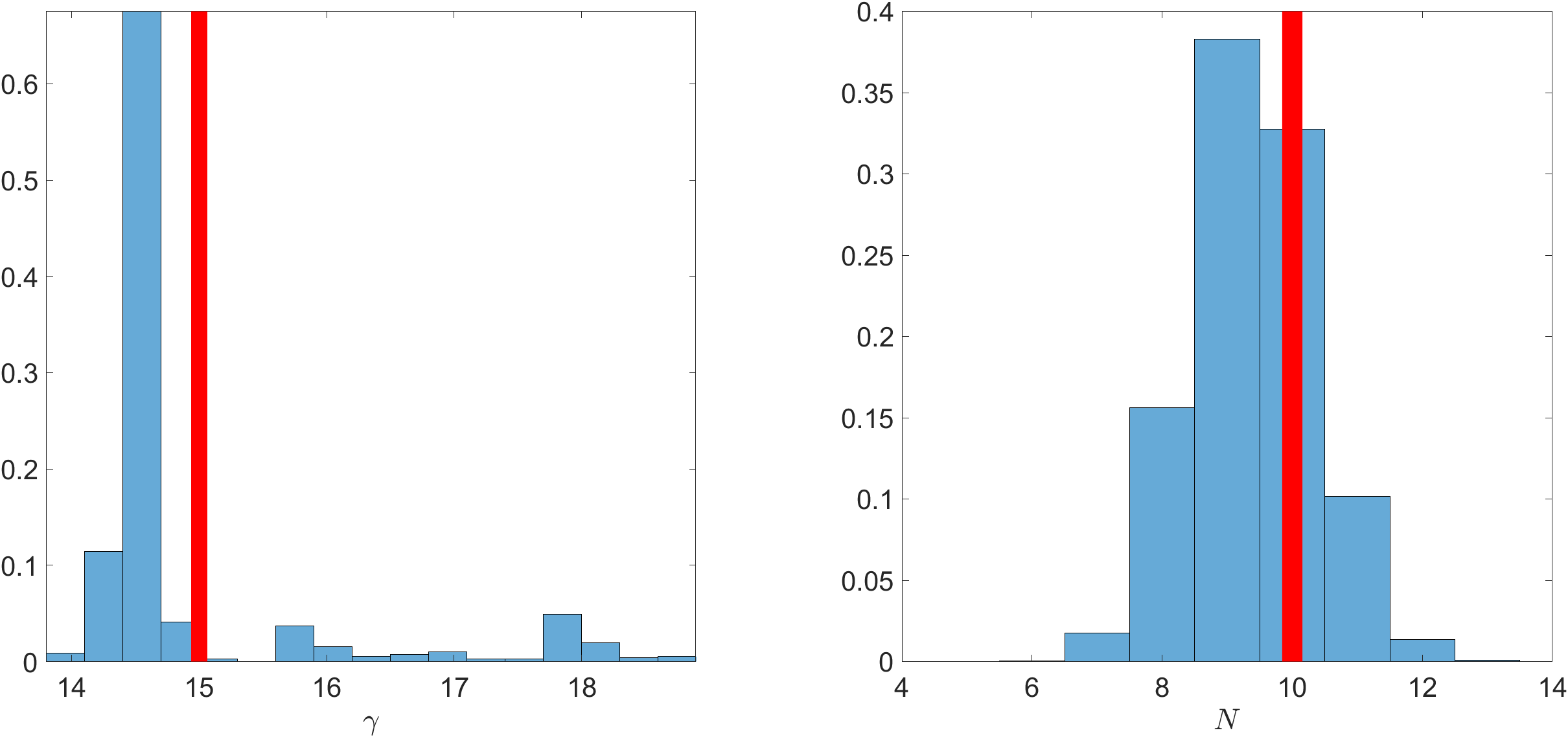}}
    \includegraphics[width=\textwidth]{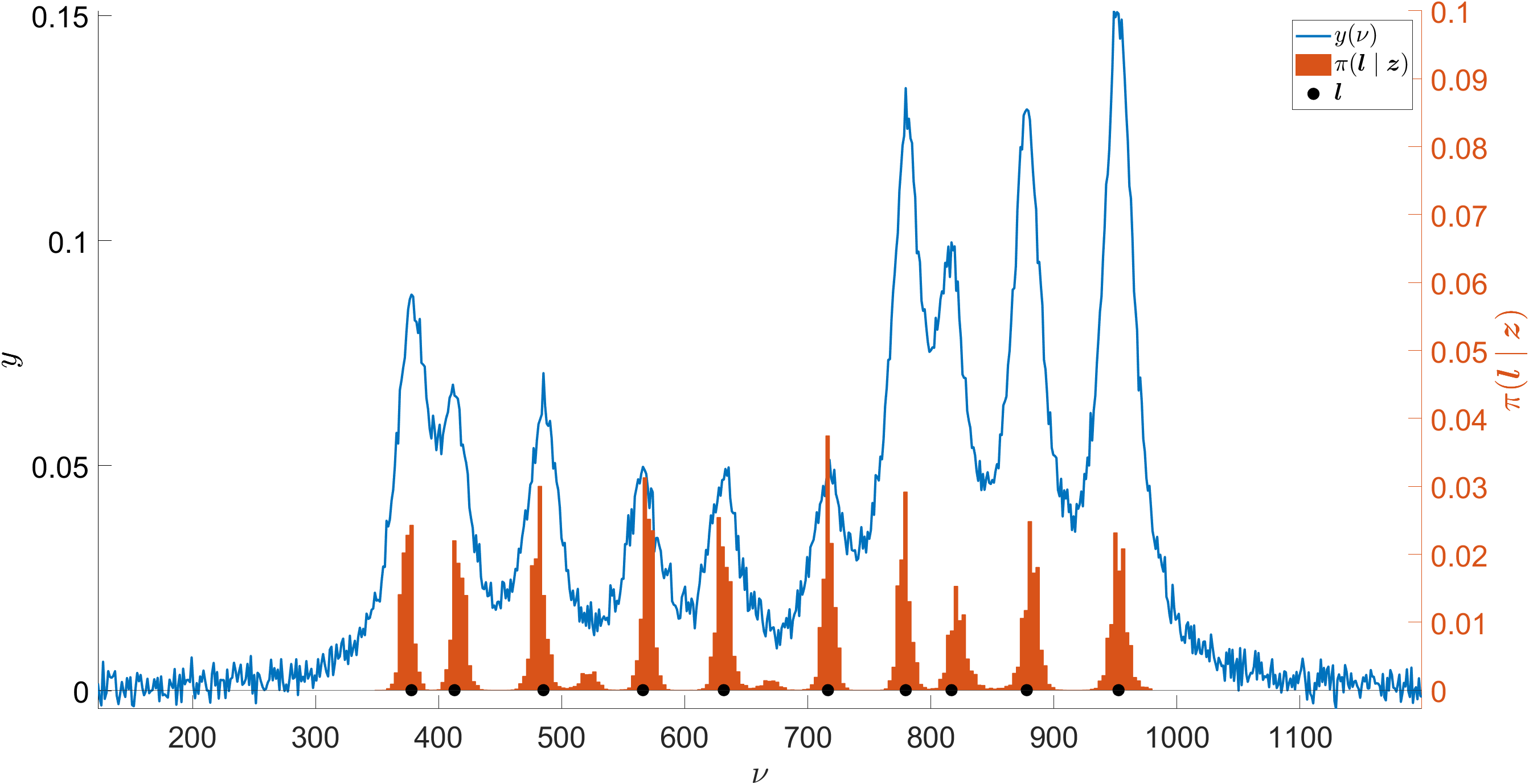}
    \caption{At the top, posterior distributions for the line shape parameter $\gamma$ and the number of line shapes $N$ along with their respective true parameter values used to generate the synthetic spectrum in red. At the bottom, the corresponding synthetic spectrum (in blue) and the corresponding location posterior $\pi( \bm{l} \mid \bm{z})$ (in red). Blacks dots denote the locations used to generate the spectrum.}
    \label{im:posteriorDataLGCP_voigtSameG}
\end{figure}

The posterior distributions for the peak locations for a synthetic spectrum used in the SBC study are shown in Figure~\ref{im:posteriorDataLGCP_voigtSameG}.
We can see here that the true peak locations $\bm{l}$ are contained within the posterior distribution $\pi( \bm{l} \mid \bm{z})$ constructed with the sampled local maxima of the LGCP.
Figure \ref{im:posteriorDataLGCP_voigtSameG} also shows the posterior distributions for the Lorentz line width parameter $\gamma$ along with the posterior for the number of peaks $N$.
Both posteriors contain the true parameter values used to generate the spectrum.
We also calculated the average bias of $-0.1$ for the posterior mean estimates of the number of peaks $N$, as well as root-mean square error (RMSE) of 1.52 and coverage of 0.97 for the 95\% posterior credible intervals in the SBC study.

\begin{figure}
 {\hspace{-0.8cm} \includegraphics[width=0.91\textwidth]{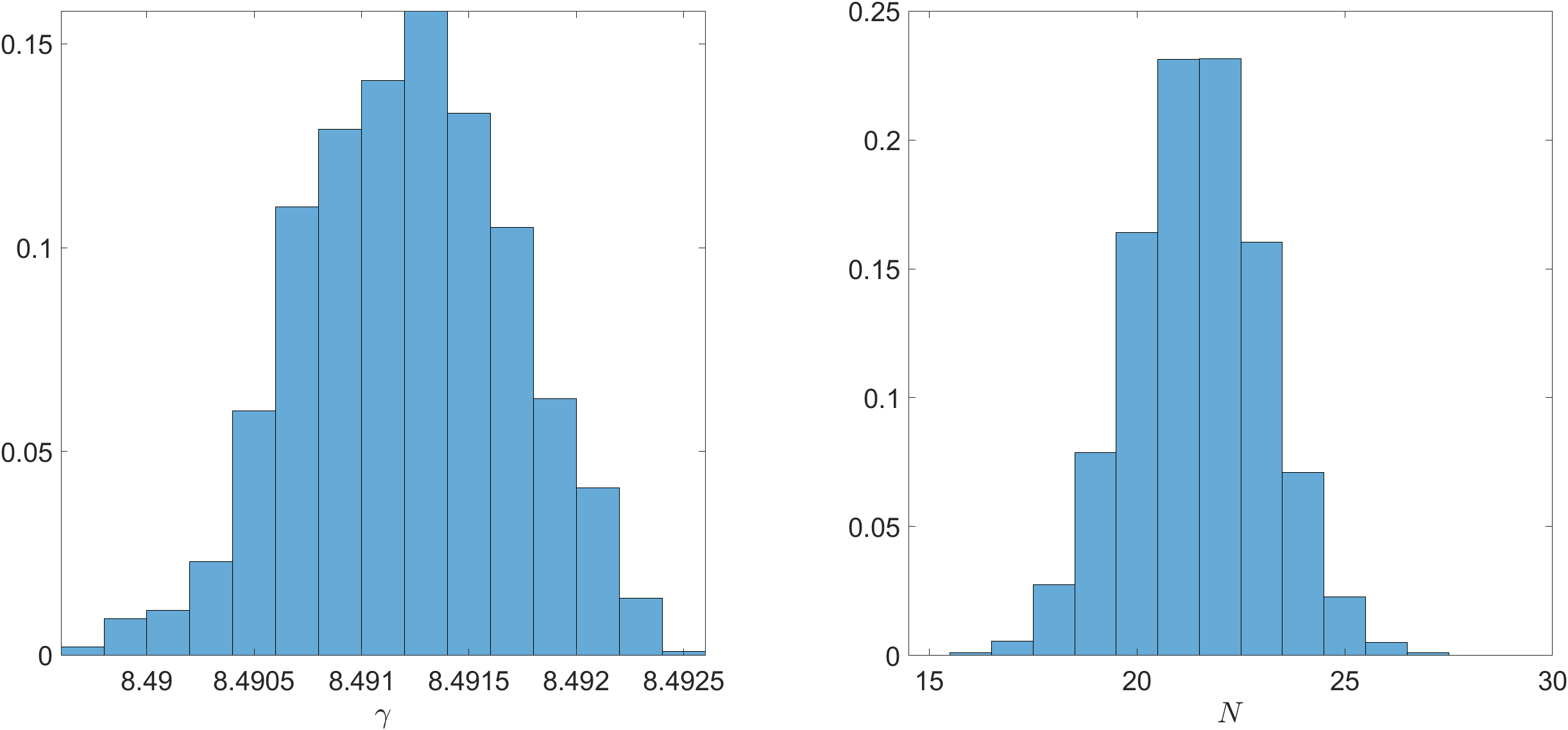}}
    \includegraphics[width=\textwidth]{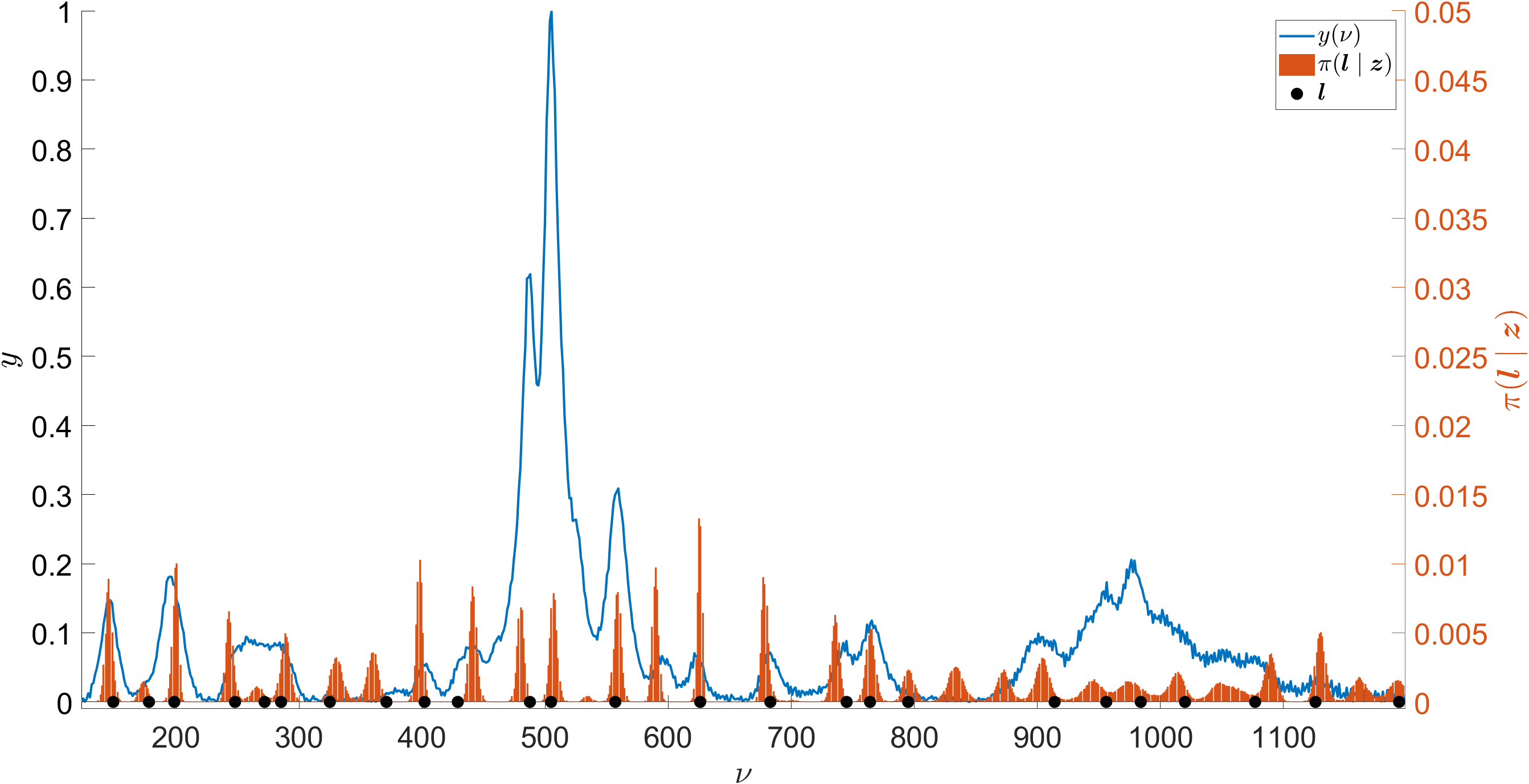}
    \caption{At the top, posterior distributions for the line shape parameter $\gamma$ and the number of line shapes $N$. At the bottom, the observed Raman spectrum of anorthite (in blue) and the corresponding location posterior $\pi( \bm{l} \mid \bm{z})$ (in red). Blacks dots denote the peak locations found in literature \cite{Freeman:2008}.}
    \label{im:posteriorDataLGCP_experimentalRaman}
\end{figure}
We analyze a Raman spectrum of anorthite (Ca Al$_2$ Si$_2$ O$_8$), a type of feldspar, which was obtained from the RRUFF database \cite[ID: R040059]{RRUFF}.
This sample is from the collection of the University of Arizona Mineral Museum.
We use the same prior distributions for all of the experimental data as for the synthetic spectra, detailed in Table~\ref{tb:priors}.
The posterior distributions for the line-shape parameter $\gamma$ and number of line shapes $N$ are shown in Figure \ref{im:posteriorDataLGCP_experimentalRaman}.
The 25 known peak locations for anorthite are given in Table 4 of \cite[p. 1488]{Freeman:2008}. These locations are illustrated with black dots in Figure \ref{im:posteriorDataLGCP_experimentalRaman}, along with the estimated  posterior distributions $\pi( \bm{l} \mid \bm{z})$.
Our method has detected some additional peak locations which appear reasonable, including one near 600cm$^{-1}$ that corresponds with a clear hump in the data, but which seems to have been previously unknown.

\section{Conclusions}
\label{sec:conclusions}
We present a Bayesian model for line narrowing of spectroscopic data that is applicable for any parameterizable kernel function.
In this paper, we particularly focus on Lorentzian and Voigt line shapes, which are typical of electromagnetic spectra.
The key innovation of our method is the use of a log-Gaussian Cox process to provide an interpretable posterior for the line narrowing and correct for unwanted peak splitting effects due to Fourier self-deconvolution.
This addresses the major limitations of the LOMEP algorithm for line narrowing \cite{Kauppinen:91,Kauppinen:1992}.

In many real-world applications, the true peak locations are unknown {\em a priori}, which limits the application of existing methods such as \cite{Moores:2018,Ritter:1994}.
Our proposed method provides posterior distributions for the peak locations, along with the line-shape parameters.
These can then be used as input for further chemometric analysis.
We have validated our proposed method using synthetic data sets and simulation-based calibration, demonstrating that the smoothed LGCP posterior is able to recover the true peak locations.
We also applied our method to three experimental Raman spectra, one of which that exhibited a low signal-to-noise ratio.
In all three cases, were able to obtain an interpretable posterior for the peak locations that corresponded well with the known spectroscopic properties of the particular mineral sample and the protein samples.
This is in contrast to many existing methods, such as \cite{Frohling:2016,Razul:2003}, that only work well in low-noise environments, or with a small number of peaks.

Although we have focused mainly on Raman spectroscopy here, our method is much more broadly applicable.
It has the potential to be used for practically any spectroscopic measurement of electromagnetic phenomena, from X-rays \cite{suuronen:2020} to radar \cite{virtanen:2021}.
By substituting the Lorentzian or Voigt line shapes with a suitable alternative, it could be applied to any time series where the periodogram can be represented as a convex combination of known spectral density functions, such as measurements of ocean waves \cite{Stewart:2004}.
Depending on the application, a different choice of point process model might also be needed.
For example, peak locations in mass spectrometry might be better modelled as a self-exciting point process, such as a Hawkes process \cite{Rasmussen:2011}.
%
%
\section*{Acknowledgments} This work has been funded by the Academy of
Finland (project numbers 327734, 334816, and 336787). The authors thank Andreas Rupp for helpful conversations during the preparation of this manuscript. We also thank the associate editor and two reviewers for their thoughtful comments and suggestions.

\clearpage
\bibliographystyle{./aims/AIMS}
\bibliography{ref}

\medskip
Received xxxx 20xx; revised xxxx 20xx.
\medskip

\end{document}